\documentclass[twocolumn,twoside,slac_two]{revtex4}
\usepackage{graphicx}
\usepackage{fancyhdr}
\usepackage{hyperref}
\begin{document}

\title{The discovery of $\gamma$-ray emission from Nova Sco 2012:\\An analysis using reprocessed Pass7 data}

%

\author{A. B. Hill}
\affiliation{W. W. Hansen Experimental Physics Laboratory, KIPAC, Department of Physics and SLAC National Accelerator Laboratory, Stanford University, Stanford, CA 94305, USA }
\affiliation{Faculty of Physical \& Applied Sciences, University of Southampton, Southampton, SO17 1BJ, UK }
\author{on behalf of the Fermi Large Area Telescope Collaboration}

\begin{abstract}
In March 2010 the Large Area Telescope on-board the \textit{Fermi Gamma-ray Space Telescope} discovered for the first time $>$100 MeV $\gamma$-ray emission from a nova within our galaxy, V407 Cyg. ÊThe high-energy spectrum and light curve was explained as a consequence of shock acceleration in the nova shell as it interacts with the local ambient medium.Ê It was suspected that the necessary conditions for high-energy emission from novae would be rare.Ê In June 2012 the LAT detected a new flaring source, Fermi J1750$-$3243, which is spatially coincident and contemporaneous with a new nova, Nova Sco 2012. ÊWe report on the exciting discovery of this new,  Ô$\gamma$-rayÕ nova and present a detailed analysis of its high-energy properties.
\end{abstract}

\maketitle

\thispagestyle{fancy}


\section{V407 Cyg: The first $\gamma$-ray nova}
On 11 March 2010 Japanese amateur astronomers reported the discovery of a new 8$^{th}$ magnitude nova in the Cygnus constellation \citep{ref1}.  The nova was identified as originating from the known symbiotic binary, V407 Cyg.  Symbiotic binaries are systems comprised of a red giant star and a hot white dwarf which is typically accreting material from the red giant via its stellar wind or Roche-lobe overflow.  In the case of V407 Cyg it hosts a Mira-type variable red giant and so the white dwarf is embedded in a particularly dusty environment generated by the heavy wind of the donor star.

\begin{figure}[h,b,t]
\centering
\includegraphics[width=80mm]{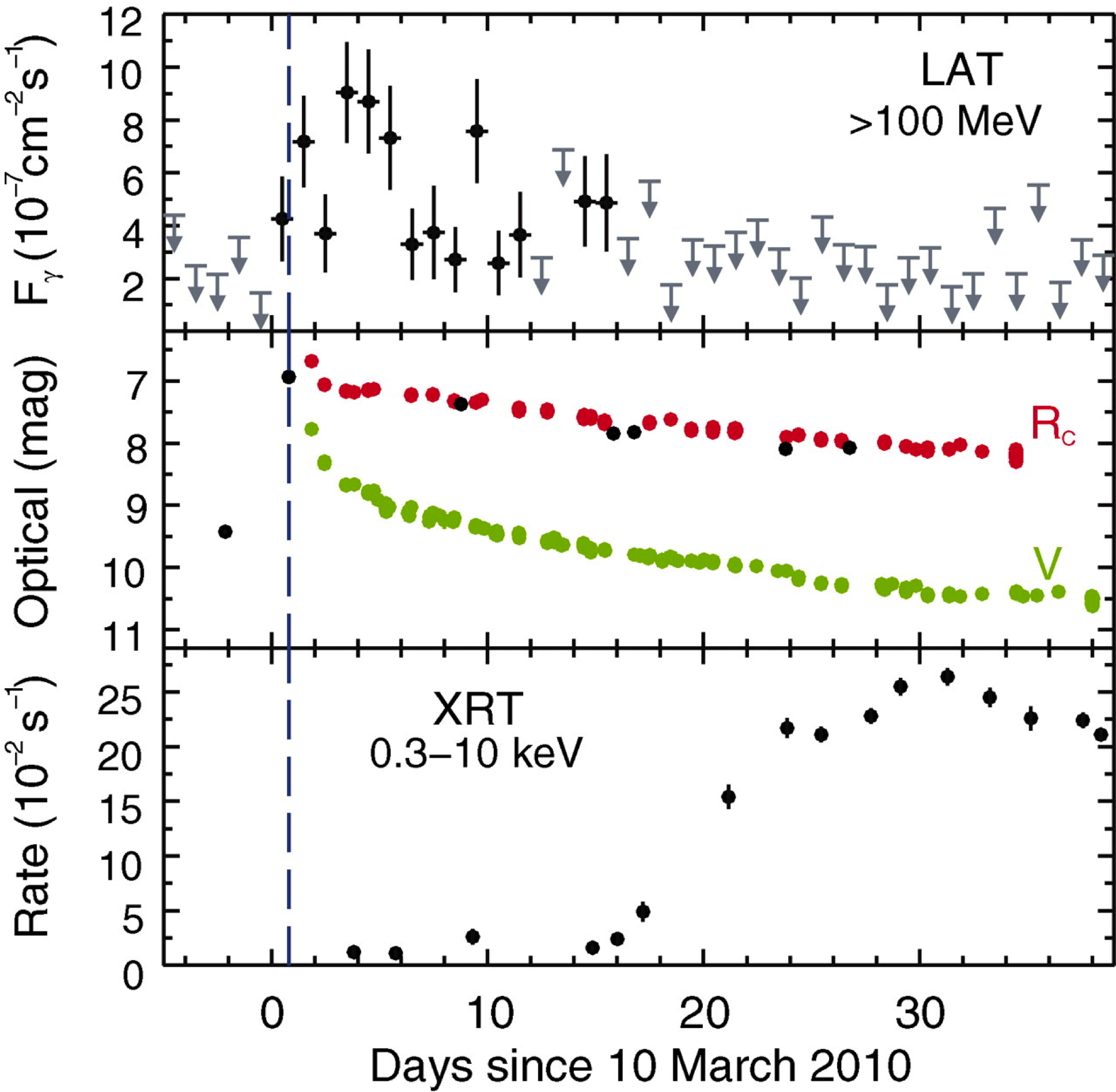}
\caption{The multi-wavelength light curve of the nova eruption in V407 Cyg in March 2010 \citep{ref3}.}\label{FigV407Cyg}
\end{figure}

The discovery of a classical nova event in this system was completely unexpected.  A further unexpected discovery came when the \textit{Fermi} Large Area Telescope (LAT) announced a detection of $\gamma$-ray emission above 100 MeV from the nova \citep{ref2,ref3}. The $\gamma$ rays were detectable for approximately two weeks after the optical nova onset with a peak flux above 100 MeV of $9 \times 10^{-7}$ ph cm$^{-2}$ s$^{-1}$.  The onset of $\gamma$-ray emission was consistent with the optical onset.  The multi-wavelength light curves of the 2010 V407 Cyg eruption are shown in Figure~\ref{FigV407Cyg}.  The features of these light curves were explained to be a consequence of the nova ejecta colliding with the red giant stellar wind and forming a shock where particles could be accelerated to high energies and consequently produce $\gamma$ rays via leptonic and/or hadronic processes.

\begin{figure*}[t]
\centering
\includegraphics[width=175mm]{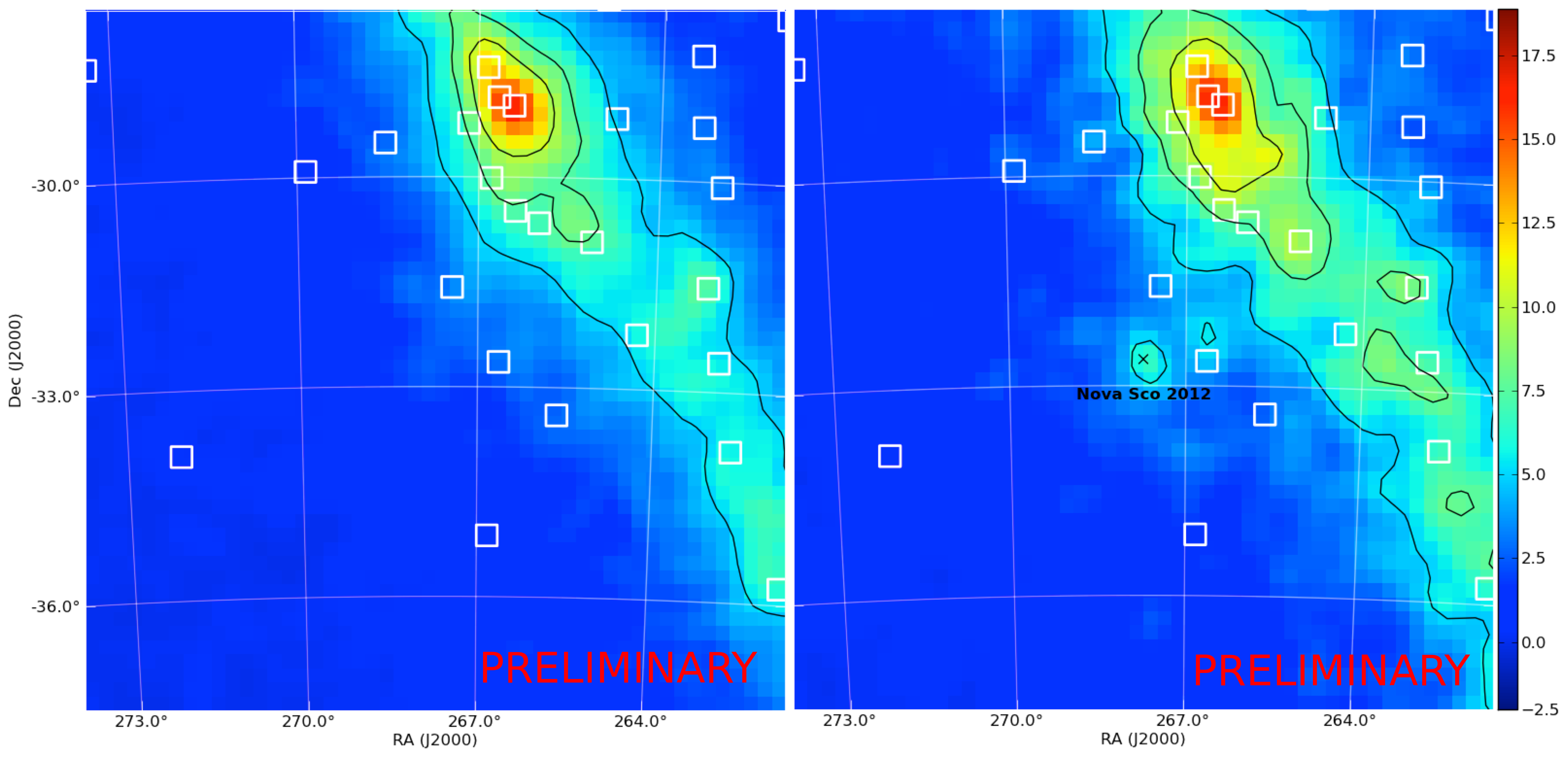}
\caption{Smoothed LAT counts maps of the region around Fermi J1750$-$3243; white squares indicate the location of 2FGL catalog sources and the black cross the known position of Nova Sco 2012.
\emph{Left}: The month from mid-May 2012, preceding the $\gamma$-ray flare. \emph{Right}: The same region for two weeks after the onset of $\gamma$-ray activity.  Contours indicate regions of 5, 7.5 \& 10 counts.  N.B. the counts have been rescaled as if both maps had equal exposure.
} \label{FigCtsMap}
\end{figure*}

\section{A new Galactic transient: \\Fermi J1750$-$3243}
In June 2012 the Routine Science Processing (RSP) pipeline reported a potential week long flare from one of the monitored X-ray binaries; for details of the RSP binaries monitoring program see \url{https://confluence.slac.stanford.edu/download/attachments/102860834/2011Glanzman_v2.pptx}.  

A detailed follow-up analysis identified a new $\gamma$-ray source, Fermi J1750$-$3243, that was not consistent with any of the known 2FGL catalog sources \citep{ref5}.  This new source was localized to RA = 267.727$^\circ$, Dec = $-$32.720$^\circ$ with a 95\% error radius of 0.122$^\circ$ \citep{ref4}. The source was not consistent with the location of the monitored X-ray binary, it was something totally new.  Figure~\ref{FigCtsMap} shows LAT counts map of the region around Fermi J1750$-$3243 for the month preceding the onset of the transient and the $\sim$two week period of activity.  A clear excess of counts is evident at a location inconsistent with any of the 2FGL catalog sources.

The location of the new LAT transient was consistent with the report of a newly discovered optical nova, MOA 2012 BLG$-$320 (Nova Sco 2012) which had entered into optical outburst between June 1.77--2.15 2012 when it brightened dramatically in the I band from 17$^{th}$ magnitude to 11$^{th}$  magnitude \citep{ref6}. Subsequent IR spectral observations on June 17.879 indicated that that it was an Fe-II nova event with an ejecta velocity of $\sim$2,200 km s$^{-1}$ \citep{ref7}.  It appeared that \textit{Fermi} had discovered another `$\gamma$-ray nova'.

\section{Data Analysis}
With more than four years of in-flight data in hand the \textit{Fermi} LAT collaboration has been assessing some of the primary sources of systematic uncertainties  and has traced some of them back to the use of non-optimal calibration constants for some of the detectors.  The collaboration has been updating these constants and used them to reprocess the raw data to improve the source analyses\footnote{The Pass 7 reprocessed data is currently available for internal analysis and assessment within the collaboration and will be made public by the end of 2013}. For further details of the Pass 7 reprocessed data see \citet{ref8, Pass7REP}

\subsection{Reprocessed Pass 7 data}
Some of the key improvements of the Pass 7 reprocessed data, also known as P7REP, include:
\begin{itemize}
\item An improved PSF above 3 GeV.
\item Change in background contamination of event classes.
\item A stabilized energy scale.
\end{itemize}
Assessments of the impact of the data reprocessing have shown that $\sim$25\% of events have been re-assigned to a different event class. A variety of sources have been re-analysed and it has been found that P7REP data are of better quality and have lower systematic uncertainties than the original P7 data indicating that the overall improvement in data quality has lead to better standard source analysis \citep{Pass7REP}.

\begin{figure}[b,t]
\centering
\includegraphics[width=80mm]{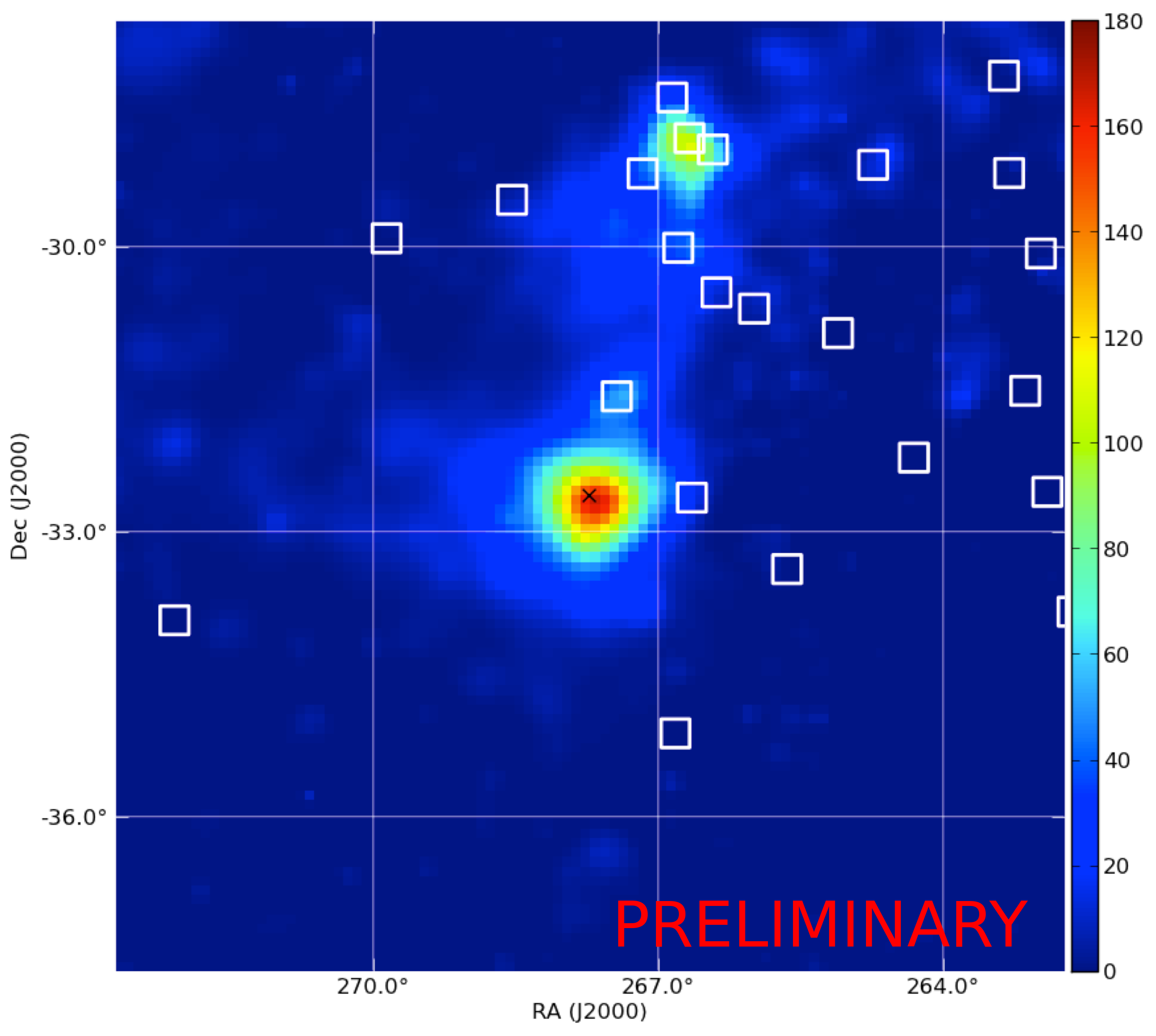}
\caption{The TS map of the region when only modelling the diffuse background components.  An additional source at a location inconsistent with any 2FGL catalog source is clearly evident.
}\label{FigTSMap}
\end{figure}

\subsection{Preliminary LAT analysis}
Having identified a new Galactic transient consistent with a new classical nova, Nova Sco 2012, we preceded to analyze the source behaviour with the new P7REP  data using the standard likelihood analysis tools included in the \textit{Fermi} \textsc{Science Tools}.  Some of the analysis details including event selection are listed below:

\begin{itemize}
\item Time range: 15.0 June -- 3.0 July 2012
\item Energy range: 0.06 -- 300 GeV
\item IRFs: P7\_V6MC; the Monte-Carlo IRFs were the best available at the time of analysis \citep[see][for further information]{Pass7REP}
\item Galactic and isotropic diffuse templates derived from the original Pass 7 data used (may underestimate the flux by $\sim$20\%; new models were not available at the time of this analysis)
\end{itemize}

\subsection{Results}

\begin{figure}[t, b]
\centering
\includegraphics[width=80mm]{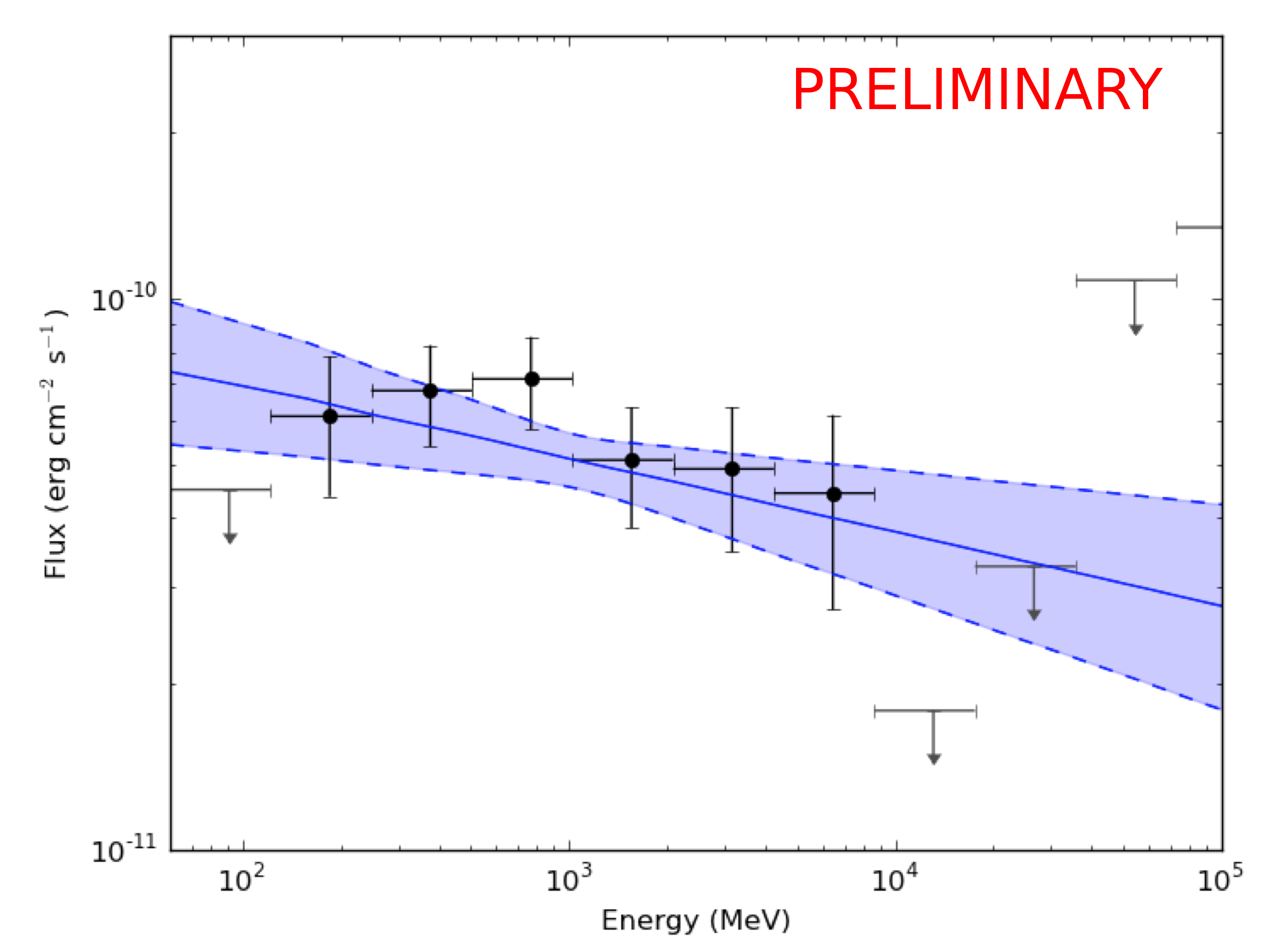}
\caption{The best-fit power law spectrum from the LAT data;  the black points have a TS$>$9 while the grey arrows indicate 95\% flux upper limits.  The shaded blue region indicates the 2$\sigma$ boundary inferred from the fitting errors.
}\label{FigScoSpec}
\end{figure}

The source is detected clearly with a significance of $>$12$\sigma$ (TS $=$ 181) and is definitely not associated with any previously detected LAT source.  Figure~\ref{FigTSMap} shows the TS map of the region when only the diffuse background components are modelled, the nova is evident in the centre of the field.   The LAT spectrum is best characterized by a simple power law model:
\begin{eqnarray}
F(E) & = & kE^{-\Gamma}
\end{eqnarray}
\noindent with an index of $\Gamma$ = 2.13 $\pm$ 0.04 and an integrated flux above 60 MeV of  $(9.6 \pm 0.7) \times 10^{-7}$ ph cm$^{-2}$ s$^{-1}$.  Figure~\ref{FigScoSpec} shows the LAT data points together with the best-fit power law spectrum.  Fitting the data with an exponentially cut-off power law (the best-fit spectral model for V407 Cyg) results in a slightly improved fit, $\Delta$TS$=11$, however it is not a significant enough improvement to adopt the more complex spectral model.  Significant emission is seen from $\sim$0.1--10 GeV and there is no detection of spectral variability over the duration of the flare. 

Fixing the spectral shape parameters the flux is calculated in daily bins to generate a $\gamma$-ray light curve for the source and is shown in Figure~\ref{FigScoLC}.  Approximately $\sim$14 days after the reported onset of the classical nova the $\gamma$-ray emission commences and is detectable in 10 independent daily bins starting over a total period of $\sim$12 days.

\section{Multi-wavelength results}
\begin{figure}[t,b]
\centering
\includegraphics[width=80mm]{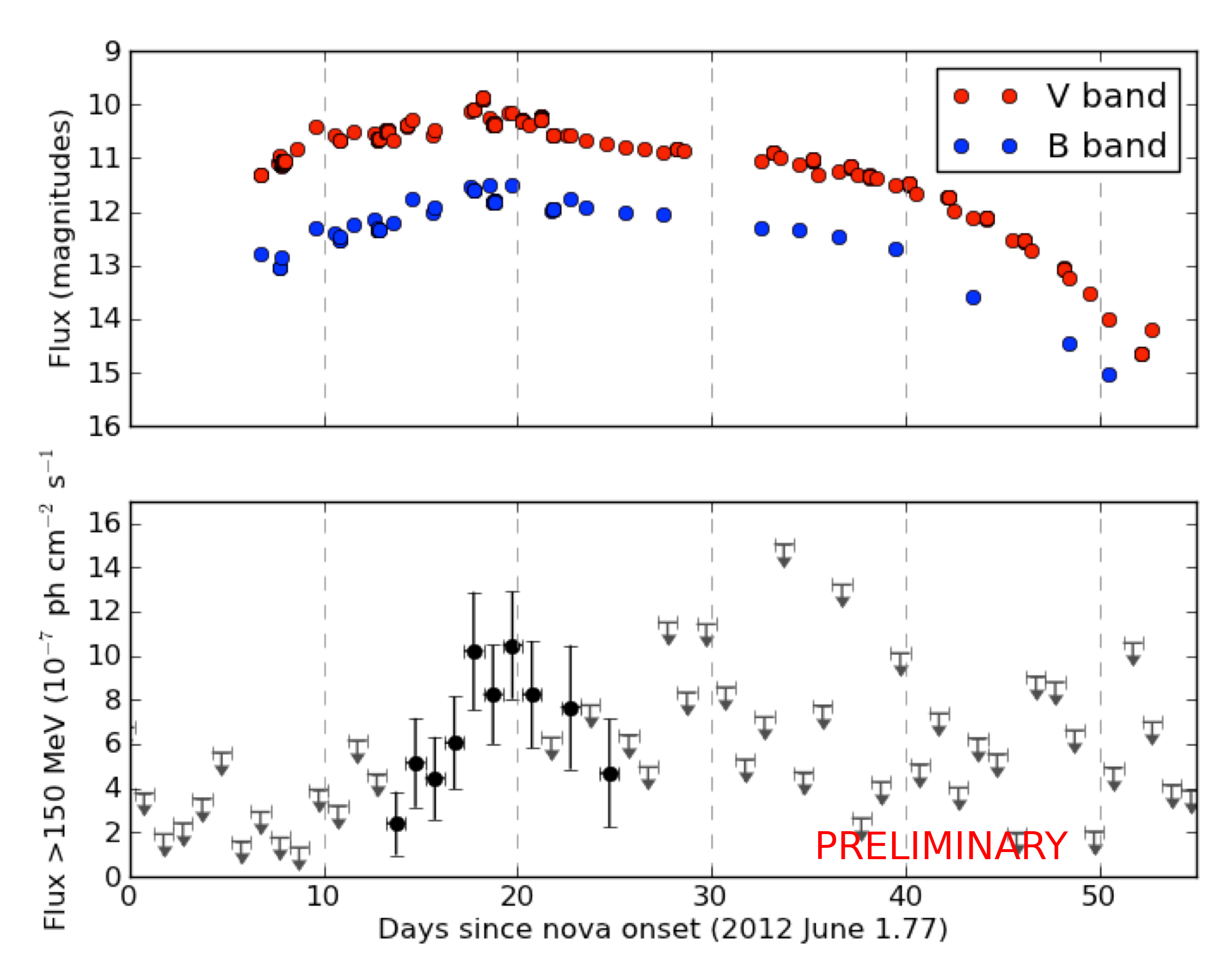}
\caption{Evolution of the nova outburst in the optical (taken from AAVSO) and the LAT $\gamma$-ray light curve. LAT flux points are shown in black with their 1$\sigma$ errors, grey arrows indicate 95\% flux upper limits.}\label{FigScoLC}
\end{figure}

Taking the optical observations from the American Association of Variable Star Observers (AAVSO; \url{http://www.aavso.org}) and plotting it with the LAT light curve it is clear that Nova Sco 2012 behaves very differently to V407 Cyg; compare Figures~\ref{FigV407Cyg}~\&~\ref{FigScoLC}. A summary of some of the different properties of these two novae events are listed in Table~\ref{pop}.  In the case of Nova Sco 2012, it very gradually reaches its optical peak slowly, $\sim$14 days after the initial rapid brightening whereas V407 Cyg reached it's maximum optical brightness almost immediately. In V407 Cyg the $\gamma$-rays were detectable at the onset of the optical eruption, however, with Nova Sco 2012 there was a delay of $\sim$14 days between the optical onset and the detection of $\gamma$-ray activity.

The decay time of the Nova Sco 2012 optical light curve places it in the fast/moderately fast novae speed class and is 4--5 times slower than V407 Cyg indicating that Nova Sco 2012 likely hosts a less massive white dwarf. Using the Maximum Magnitude rate of decline (MMRD) relation of \citet{mmrd} and the intrinsic novae colour relations of \citet{ebv} suggests that the source is 4--5 kpc distant.  Our derived value of the extinction towards the source is compatible with estimates of extinction along the line of sight, however these estimates are known to be unreliable within 5$^{\circ}$ of the Galactic plane.

Radio emission has also been reported from Nova Sco 2012 however X-ray observations with Swift--XRT have to date not detected anything \citep{RadioATel,SwiftATel}.  

\begin{table*}\caption{A comparison of some of the properties of the 2010 classical nova event in V407 Cyg and Nova Sco 2012.}
\begin{center}
\begin{tabular}{lcc}
\hline
System & Nova Sco 2012 & V407 Cyg   \\
\hline
Optical companion & ? & Mira-type red giant \\
Distance & $\sim$4--5 kpc & 2.7 kpc\\
Nova spectral class & Fe-II & He/N \\
Speed class & Fast/moderately fast & Very fast\\
& t$_{2} \sim$25 days & t$_{2} \sim$5.9 days \\
$\gamma$-ray spectrum & Power law & Exponentially cutoff power law\\
$\gamma$-ray duration & $\sim$12 days & $\sim$16 days \\
Optical/$\gamma$-ray delay & $\sim$14 days & $<$3 days \\
Average $\gamma$-ray flux &  9.6$ \times 10^{-7}$ ph cm$^{-2}$ s$^{-1}$ & 4.4$ \times 10^{-7}$ ph cm$^{-2}$ s$^{-1}$\\
\hline
\end{tabular}
\end{center}
\label{pop}
\end{table*}

\section{Summary}
We have discovered a new Ô$\gamma$-rayÕ nova, the second member of a new high-energy source population.  It has many similar $\gamma$-ray properties to the first LAT detected nova, V407 Cyg: spectral shape; duration; peak flux.  Conversely, the optical properties are in stark contrast showing a much slower nova, potentially $\sim$1.6 times more distant and with an unidentified optical companion.  The relationship between the optical and $\gamma$-ray properties is also not clear: in the case of V407 Cyg the high-energy and optical onset were compatible with being simultaneous in the case of Nova Sco 2012 the $\gamma$-rays lag the optical onset by $\sim$14 days although they do peak at the same time at both wavelengths.

The $\gamma$-ray production mechanism invoked to explain V407 Cyg was particle acceleration at the shock front produced by the collision of the nova ejecta with the heavy stellar wind of its red giant companion.  Environments such as these are very rare in nova systems and while the donor star in the Nova Sco 2012 system is currently unknown it is not unreasonable to expect it to belong to the more typical population of low-mass dwarf donors.  This would require a different model for the $\gamma$-ray production as such a system would not have a heavy stellar wind.

The potential appears to exist for novae, in general, to be capable of producing $\gamma$-rays, however, the mechanism behind this emission is still to be understood and has the potential to be different within the novae sub-classes.

\bigskip 
\begin{acknowledgments}
A.~B. Hill acknowledges that this research was supported by a Marie Curie International Outgoing Fellowship within the 7th European Community Framework Programme (FP7/2007--2013) 
under grant agreement no. 275861. 

The \textit{Fermi} LAT Collaboration acknowledges generous ongoing support
from a number of agencies and institutes that have supported both the
development and the operation of the LAT as well as scientific data analysis.
These include the National Aeronautics and Space Administration and the
Department of Energy in the United States, the Commissariat \`a l'Energie Atomique
and the Centre National de la Recherche Scientifique / Institut National de Physique
Nucl\'eaire et de Physique des Particules in France, the Agenzia Spaziale Italiana
and the Istituto Nazionale di Fisica Nucleare in Italy, the Ministry of Education,
Culture, Sports, Science and Technology (MEXT), High Energy Accelerator Research
Organization (KEK) and Japan Aerospace Exploration Agency (JAXA) in Japan, and
the K.~A.~Wallenberg Foundation, the Swedish Research Council and the
Swedish National Space Board in Sweden.  Additional support for science analysis during the operations phase is gratefully
acknowledged from the Istituto Nazionale di Astrofisica in Italy and the Centre National d'\'Etudes Spatiales in France.

Work supported by Department of Energy contract DE-AC03-76SF00515.
\end{acknowledgments}

\bigskip 

\end{document}